# Designing a Social Media Analytics Dashboard for Government Agency Crisis Communications

## Research-in-progress

### Ali Sercan Basyurt
Department of Computer Science and Applied Cognitive Science
University of Duisburg-Essen
Duisburg, Germany
Email: ali-sercan.basyurt@uni-due.de

### Julian Marx
Department of Computer Science and Applied Cognitive Science
University of Duisburg-Essen
Duisburg, Germany
Email: julian.marx@uni-due.de

### Stefan Stieglitz
Department of Computer Science and Applied Cognitive Science
University of Duisburg-Essen
Duisburg, Germany
Email: stefan.stieglitz@uni-due.de

### Milad Mirbabaie
Department of Information Systems
Paderborn University
Paderborn, Germany
Email: milad.mirbabaie@upb.de

## Abstract

Social media have become a valuable source for extracting data about societal crises and an important outlet to disseminate official information. Government agencies are increasingly turning to social media to use it as a mouthpiece in times of crisis. Gaining intelligence through social media analytics, however, remains a challenge for government agencies, e.g. due to a lack of training and instruments. To mitigate this shortcoming, government agencies need tools that support them in analysing social media data for the public good. This paper presents a design science research approach that guides the development of a social media analytics dashboard for a regional government agency. Preliminary results from a workshop and the resulting design of a first prototype are reported. A user-friendly and responsive design that is secure, flexible, and quick in use could identified as requirements, as well as information display of regional discussion statistics, sentiment, and emerging topics.

# 1    Introduction

Social media have become an important part of the life of billions and influences the way people consume and share information (Mirbabaie et al. 2021). Theoretical approaches and methods for analysing social media data are especially relevant for the information systems (IS) discipline. One important methodological standpoint is taken by the Social Media Analytics (SMA) approach (Stieglitz et al. 2018). As a result of highly relevant phenomena revolving around social media and the massive amounts of data generated every day, SMA has gained popularity among IS scholars (Choi et al. 2020). It can be understood as *"an emerging interdisciplinary research field that aims on combining, extending, and adapting methods for analysis of social media data"* (Stieglitz et al. 2014). Important contributions to SMA have been made by many disciplines, including IS. SMA deploys structured, semi-structured, and unstructured social media data (e.g. text, images, time-series data, metadata, network relationships), and enables researchers or organisations to interpret public social media data for assessing people's behaviours and attitudes (Kurniawati et al. 2013). Therefore, SMA is highly relevant in various contexts, e.g. political communication, crisis management, or business (Choi et al. 2020). For this purpose, application-driven research has suggested dashboard interfaces as suitable tools for handling social media data (Avvenuti et al. 2018).

In this paper, we focus on the application domain of government agency (GA) communications. Recent events of societal impact such as the Australian Bushfires, the COVID-19 pandemic, or the flooding in Western Germany show the importance of GA to be able to understand how information diffuses through social media or how predictions about future behaviour can be made based on social media data. The main problems of application driven IS research in this specific context are threefold. First, the opaque fields of jurisdiction in terms of SMA inside GA can lead to differential requirements for SMA artefacts. Second, the differing levels of technical savvy inside and across GA make it difficult to determine the average end user. Third, due to different sizes and areas of responsibilities, the challenges, and opportunities of SMA artefacts may significantly differ among GA, for example between a state-wide administration and a regional municipality. Regional GA are regularly confronted with different crises spanning from political crises to natural disasters which emphasizes their need for training and instruments to handle these crises. The purpose of this study is to understand the SMA requirements of regional GA and to provide a dashboard prototype that addresses the collected requirements tailored to this type of GA. Hence, our research is guided by the following research question:

**RQ1:** *What requirements do regional GA have for a Twitter-based SMA dashboard?*

**RQ2:** *How can regional GA requirements be integrated in a Twitter-based SMA dashboard design?*

In order to conduct our research, we employ the Design Science Research (DSR) approach (Hevner et al. 2004). We follow the sequence of six steps as described by Peffers et al. (2007) to structure our approach. This research-in-progress paper presents the first three steps of the sequence: (1) problem identification and motivation, (2) objectives of a solution, and (3) design and development. This work will contribute to practice by showcasing how institutional units such as GA with assume SMA methods and instruments and how SMA can be utilized for the common good. This DSR project will shed light on how GA can improve their knowledge about actors and dynamics of social media communication and how to eventually include social media as a barometer of public participation in their decision-making. This paper's contribution to knowledge is directed at the experiences we make while collaborating with GA and how institutional structures and power influence the DSR process.

The remainder of this paper is structured as follows. First, we present related work in section 2. Subsequently, the research design and DSR project information are outlined in section 3. Preliminary findings are presented in section 4-5 and next steps and concluding remarks are provided in section 6.

# 2    Related Work

## 2.1    SMA for Government Agency Crisis Management and Communication

Social media are self-governed information systems that can be utilised with immense spatial and temporal flexibility (Mirbabaie et al. 2020). A key capacity of social media, that is instant information dissemination and extraction, can be of utmost value to organisations assuming a leading role in crisis management such as GA. For purposes of information *extraction*, GA can use SMA to keep track of public discussions revolving around an acute crisis (Chen et al. 2020, Guo et al. 2021). Moreover, SMA can be used to inform decision-making concerning the deployment of resources, e.g. in case of social media users reporting about circumstances in certain local areas (Marx et al. 2018). Information *dissemination*, in turn, deals with the capability of GA to actively exercise influence on public discussions by providing relevant intelligence. Here, SMA can inform GA about the most pressing information gaps among the public, prevalent narratives, and most influential social media users engaging in the communication.

IS scholarship on social media crisis communication, in this regard, has spawned two different research streams to contribute to this domain. First, the last decade has seen several retrospective case studies analysing social media data from a variety of crises, offering deeper knowledge of the extraction and dissemination processes, roles, and mechanisms as they evolve in social media crisis communication. Examples for such applications of SMA include research into the Red River flood and the Oklahoma Fires in 2009, the 2010 Haiti earthquake, the 2011 Queensland floods, the 2011 Tunisian revolution, the 2011 Norway terrorist attacks, the 2011 Egypt revolution and uprisings, the Hurricane Sandy in 2012, the Boston Marathon bombing in 2013, the typhoon Haiyan in the Philippines 2013, the Sydney siege in 2014, the Paris and Brussels terror attacks in 2015/16 (Reuter and Kaufhold 2018), Hurricane Harvey in 2017 (Mirbabaie et al. 2020), and the coronavirus pandemic in 2020 and beyond (Shahi et al. 2021). Second, IS scholarship has emphasised the need for solutions to support stakeholders of social media crisis communication with real-time SMA. This research stream is primarily reflected in action research and design science research projects testing and evaluating artefacts for crisis management. Advancing knowledge in this research stream is important because social media crisis communication is time-critical and both information extraction and dissemination must happen expeditious (Gálvez-Rodríguez et al. 2019) if crisis managers aim to utilise platforms such as social media to support their decision-making. This paper follows the logic of the second research stream. To make a significant contribution to this body of knowledge, we now review some of the challenges that emerged in practically oriented studies involving SMA and organisations in the context of social media crisis communication.

## 2.2 Challenges of Social Media Analytics for Government Agencies

In recent years, communication on social media has had a massive impact on society, and particularly on citizen's decision-making processes, enabling new forms of public discourse [6] and challenging societal cohesion. Such challenges have been highlighted by GA such as the Council of Europe (Wardle and Derakhshan 2017), or the NATO Strategic Command (NATO 2017). Social media communication plays a particularly crucial role for citizens and GA during crisis situations which pose an urgent need to rapidly find up-to-date situational information and disseminate it effectively across directly and indirectly affected populations, for example during natural disasters such as floods or human-made crises such as terror attacks (Fathi et al. 2020). Here according to Rathore et al. (2021), the (1) verification of available information from a variety of official and unofficial sources poses a particular challenge for GA. There is also a need to (2) ensure that accurate information receives wider circulation and impact than unsubstantiated rumours or deliberate disinformation. GA are confronted with an immense amount of new social media data that is continuously generated, requiring deep knowledge in social media analytics to get meaningful insights. At the same time, GA experience (3) a lack of comprehensive training and instruments. Therefore, methods and tools that are easy to use are urgently required to enable GA to engage in SMA while managing emerging risks for their field of activity. Previous research has explored the potentials of designing such methods and tools for emergency management agencies (EMA), which include police and fire departments, and other emergency services (Mirbabaie and Fromm 2019). The requirements and technical abilities of EMA, we argue, are different for GA. Whereas EMA are experienced in using information systems to extract and disseminate information, existing processes and structures can be adjusted to social media. GA, in contrast, often rely on external advice and processes are often much less digitised (Gholami et al. 2021). Moreover, EMA are often experienced in using (command and control) information systems. The challenge for them is in implementing information from self-governed systems such as social media into their workflow. GA, in relation to EMA, are novice users of information systems for crisis management. Moreover, their objectives differ insofar as they aim more towards understanding narratives (example: misinformation) or gaining access to communities (example: refugees). Therefore, this paper explores GA requirements for SMA and presents a DSR artefact that is tailored to the challenges GA face.

## 3 Design Science Research Methodology

DSR is described as a problem-focused research paradigm with the goal of addressing a specific organizational problem. It is characterized by creating an artefact to solve these predefined problems (Hevner et al. 2004). By generating new knowledge for example in form of design processes that are identified through the creation of an artefact a contribution to this research field can be made. Since the aim of our research is to implement an artefact to help GA with their communications by analysing and visualizing social media data, the DSR approach is suitable to answer the research questions. We follow the six steps defined by Peffers et al. (2007). Table 1 depicts each step and describes actions we take. This includes the steps addressed in this paper (white background) and the steps that will be addressed in future work (grey background).

| 1. Problem identification and motivation | 2. Objectives of a solution | 3. Design and development | 5. Evaluation |
|---|---|---|---|
| GA face a plethora of information and citizen inquiries, which is amplified in crisis situations. At the same time, a public discourse evolves on social media platforms such as Twitter. Automated analysis and visualisation of this data can improve GA decision-making and enhance their ability to participate in the public discourse. The amount and verification of information shared on social media (challenge 1+2), and a lack a of training and instruments (challenge 3) make it hard for GA to engage in SMA. | The DSR artefact aims to automatically analyse and visualise social media data for GA staff in form of a dashboard. This technical solution should provide: <br>• descriptive information <br>• public sentiment over time <br>• discussion frequency over time <br>• most shared opinions and narratives, e.g. in times of crisis <br> It must enable GA to quickly recognise abnormalities in public discussions, clearly organise all relevant data, and be intuitive for users. | A first version of the dashboard is designed and implemented. It is connected to the Twitter-API and has several modules that serve the objectives. | The dashboard's usability, usefulness and applicability is evaluated with members of the GA. |
| | | **4. Demonstration** | **6. Communication** |
| | | The ability of the dashboard to address the objectives is demonstrated with members a regional GA | The results are communicated to an interested audience consisting of regional and supra-regional GA. |

*Table 1: Overview of DSR process according to the six steps by Peffers et al. (2007).*

The first step of the sequence is dedicated to identifying the problem which we intend to address with our research and the motivation for solving this issue. We focus our research on the use of social media platforms by GA. To simplify this, we assume the application scenario of a crisis. Social media user data can be useful in different phases of a crisis (Fathi et al. 2020; Marx et al. 2018). The *problem* for our research was identified by an analysis of literature, which resulted in three challenges for GA. Moreover, the *motivation* for this research stemmed from a third-party funded research project, in which the research team collaborates with a regional GA with the goal of uncovering communication patterns on social media and suggesting best practices to seek and share information in crises. The potential of improving the ability of GA to comprehend the severity of a crisis earlier, manage it more efficiently as well as make better and quicker decisions, motivates the goal of this work.

The second step in the sequence is focused on defining *objectives* for our artefact that if addressed should lead to the defined problem being solved by the dashboard. To acquire information regarding the requirements that GA have for a SMA dashboard, a workshop was conducted. This workshop involved members of a regional GA, i.e. a municipality in Northern Europe, which is responsible for a region with a population of 166,000 citizens. The municipality had experience in dealing with several crisis situations, which were all characterised by collateral social media discussions. Similar to the approach described by Grobler and de Villiers (2017) in their DSR based work the workshop included different tasks, e.g. the development of personas and a fictional crisis scenario by the staff of the municipality. As part of the workshop, the participants were asked to express which requirement they have for a dashboard based on social media data to assess and manage the crisis scenario as best as possible. These requirements were then used to derive objectives for our solution. The third step of the DSR process model sequence is focused on designing and developing the artifact that solves the identified problem.

## 4    Requirements and Objectives of the SMA Dashboard

As part of our research project, a workshop was conducted with three members of the municipality over the course of two days. The aim of the workshop was to gain insight into the requirements that regional GA have for a SMA dashboard, especially when managing a crisis. As a warm-up exercise the participants were asked to brainstorm aspects for a perfect platform that supports crisis management and the active dissemination of information on social media. Based on these requirements, objectives for the development of a SMA dashboard were identified. To identify the requirements and derive the objectives, the application domain with the focus on people, organizational systems, technical systems, and problems as well as opportunities were analysed within the workshop. These areas were focused due to our goal being the identification of problems and opportunities and the three remaining areas possibly influencing GA and their problems. To acquire information on these topics systematically, different workshop techniques were applied. This includes addressing the staff members by creating personas as well as addressing the GA structures and technical systems through the creation of a use case together with the participants. Additionally, the problems and opportunities leading to the requirements for a supporting application were addressed by developing user stories. The resulting personas describe the stakeholders and future users of a potential tool and their individual roles and goals such as a communications manager with the role to provide the public with correct and secure information as soon as possible or a crisis coordinator that desires to manage resources effectively during a crisis. The use case that was chosen by the participants was a measles-outbreak. This scenario depicts the course of a crisis as well as typical processes and actions that the municipality staff and the created personas should take during the crisis. For example, the crisis manager should acquire and provide information to the communications manager who then informs the public about the crisis by contacting the media. The data was qualitatively evaluated, summarized, and categorized into topics by two researchers. Resulting requirements and objectives are shown in table 2.

| No | Requirements | Objectives |
|----|---|---|
| 1 | Connecting systems | - |
| 2 | User friendly/ basic design | Design should be user friendly & easy to use |
| 3 | Mobile version | Mobile visualisation of application |
| 4 | Automatically post info from website on social media | - |
| 5 | Chatbot that answers questions/ provides advice and information | - |
| 6 | Enough capacity | Multiple users & expandable storage capacity |
| 7 | Dashboard<br>• Top 10 communicators<br>• Main topic on social media<br>• Statistical analysis of tweets<br>• Sentiment of tweets<br>• Identification of bots | Development of a Dashboard that includes:<br>• -<br>• Top trending Tweets<br>• general descriptive information<br>• Sentiment analysis over crisis & daily<br>• - |
| 8 | Filter information (e.g. pictures)<br>• Accumulate posts about the same topic until a maximum number is reached to respond | - |
| 9 | Accessibility | - |
| 10 | Responsive design | Adaptation to the properties of the device used |
| 11 | Different languages | - |
| 12 | Security | Data should be secured |
| 13 | Website for citizens for flat hierarchy | - |
| 14 | Documentation (e.g. journals, use cases) | - |
| 15 | Maintainability/ flexibility | The application should be easy to expand |
| 16 | Quick in runtime | The application should display results quick |
| 17 | AI-based predictions | - |

*Table 2: Requirements and objectives resulting from the GA practitioner workshop*

## 5   Design and Development

With the goal of equipping GA with a dashboard that provides them with information about public (crisis) discussions on Twitter, an artifact in the form of a dashboard was designed. To collect data from Twitter such as tweets, retweets, @mentions, and likes we used a separate application that is dedicated to collecting data specifically from Twitter. This tool employs the Twitter API and predefined search terms that are then tracked in Twitter conversations such as hashtags and usernames of registered Twitter accounts. This data is visually presented in our dashboard in multiple modules containing various charts with different information. The dashboard was developed using *Node.js*, *expressjs*, *MongoDB*, and *vue*. The modules display a loading screen to ensure that the data is fully loaded and displayed before the user can perform actions on the dashboard as well as to guarantee that the dashboard is quick during its runtime by minimizing the frequency of data being loaded. This addresses requirement and objective No 16 from Table 2. For security purposes the access to the data located on a database can be limited to the IP addresses of known users on MongoDB-Atlas as a hosting platform (requirement and objective No 12). Furthermore, the development process of the dashboard included a responsive design that ensures that the dashboard adapts to the properties of the device it is accessed with, such as a mobile device (requirements and objectives No 3, No 10). It is accessible by multiple users and its storage capacity is expandable (requirement and objective No 6). The modularity of the dashboard ensures that it is easy to use and that it can be effortlessly expanded by adding new modules (requirements and objectives No 2, No 15). Figure 1 shows a screenshot of the developed dashboard.

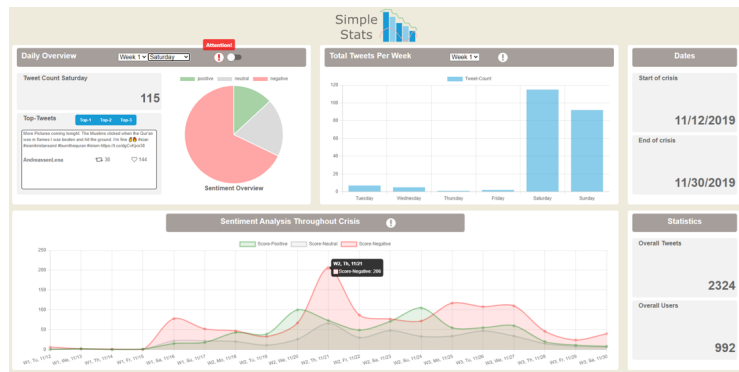

*Figure 1: Screenshot of the SMA dashboard for GA communications and crisis management*

The first module is located on the far right of the dashboard and it provides the user with an overview of the tweets included in the test dataset that is linked to the dashboard through a database. The second module is located at the bottom of the dashboard, and it is dedicated to the objective of sentiment analyses. The third module is located at the upper left corner of the dashboard that is titled daily overview. This section allows the user to select a specific day for which they want to acquire detailed information. The fourth module is located to the right of the daily overview module. This module displays the total tweets per week and therefore addresses the objective of general descriptive information about the dataset as well. These modules address requirement and objective No 7.

## 6 Discussion and Next Steps

The dashboard is focused on providing GA with an instrument that visualizes social media data to improve their crisis management capabilities. In future iterations, our goal is to further develop the artifact to address the first two challenges. Through the developed dashboard and further through its future evaluation, a contribution to research focused on the implementation of SMA for crisis management is made. By addressing several of GA needs with the artefact a practical contribution is made by providing them with an instrument that supports their work. This paper has identified challenges GA face concerning SMA. Using a DSR approach, requirements were identified with a practitioner workshop. Finally, a SMA dashboard was developed as a solution to address these requirements and objectives by presenting the desired information and following formulated conditions for the artefact as described in the discussion. Building on this work will enable GA to use SMA more effectively for communications and crisis management and inform IS research about the challenges faced by GA in employing SMA as well as how these can be overcome through DSR artefacts.

The next steps encompass the demonstration of the ability of the artifact to successfully address the problem it was developed to solve (step 4). The evaluation will involve focus group discussions with GA staff who have engaged with the artefact (step 5). Finally, the results of the project will be disseminated among the consortium of the third-party funded project and published in a scientific article (step 6) that addresses in detail the technical features of the artefact in concert with the specific challenges of GA in times of crisis.

## 7 References

Avvenuti, M., Cresci, S., Del Vigna, F., Fagni, T., and Tesconi, M. 2018. "CrisMap: A Big Data Crisis Mapping System Based on Damage Detection and Geoparsing," *Information Systems Frontiers* (20:5), Information Systems Frontiers, pp. 993–1011.

Chen Q., Min, C., Zhang, W., Wang, G., Ma, X., & Evans, R. 2020. "Unpacking the black box: How to promote citizen engagement through government social media during the COVID-19 crisis," *Computers in Human Behavior* (110:1), pp. 1-9.

Choi, J., Yoon, J., Chung, J., Coh, B. Y., and Lee, J. M. 2020. "Social Media Analytics and Business Intelligence Research: A Systematic Review," *Information Processing and Management* (57:6), Elsevier, pp. 1–18. (https://doi.org/10.1016/j.ipm.2020.102279).

Fathi, R., Thom, D., Koch, S., Ertl, T., and Fiedrich, F. 2020. "VOST: A Case Study in Voluntary Digital Participation for Collaborative Emergency Management," *Information Processing and Management* (57:4), Elsevier, pp. 1–25. (https://doi.org/10.1016/j.ipm.2019.102174).

Gálvez-Rodríguez, M. del M., Haro-de-Rosario, A., García-Tabuyo, M., and Caba-Pérez, C. 2019. "Building Online Citizen Engagement for Enhancing Emergency Management in Local European Government: The Case of the November 2015 Paris Attacks," *Online Information Review* (43:2), pp. 219–238. (https://doi.org/10.1108/OIR-09-2016-0286).

Gholami, R., Singh, N., Agrawal, P., Espinosa, K., and Bamufleh, D. 2021. "Information Technology/Systems Adoption in the Public Sector: Evidence from the Illinois Department of Transportation," *Journal of Global Information Management* (29:4), pp. 172–194.

Grobler, M., and de Villiers, C. 2017. "Shaping Solutions with a Community: The Research Design Using Design Science Research (DSR) and Case Study Research with an ICT4D Project," *CONF-IRM 2017 Proceedings*.

Guo J., Liu, N., Wu, Y., & Zhang, C. 2020. "Why do citizens participate on government social media accounts during crises? A civic voluntarism perspective," *Information & Management* (58:1), pp. 1-12.

Hevner, A. R., March, S. T., Park, J., and Ram, S. 2004. "Design Science in Information Systems Research1. Hevner, A.R., March, S.T., Park, J., Ram, S.: Design Science in Information Systems Research. MIS Q. 28, 75–105 (2004).," *MIS Quarterly* (28:1).

Kurniawati, K., Shanks, G., and Bekmamedova, N. 2013. "The Business Impact Of Social Media Analytics," in *Proceedings of the 21st European Conference on Information Systems*, pp. 1–13.

Marx, J., Mirbabaie, M., and Ehnis, C. 2018. "Sense-Giving Strategies of Media Organisations in Social Media Disaster Communication: Findings from Hurricane Harvey," in *Proceedings of the 29th Australasian Conference on Information Systems*, pp. 1–12.

Mirbabaie, M., Bunker, D., Stieglitz, S., Marx, J., and Ehnis, C. 2020. "Social Media in Times of Crisis: Learning from Hurricane Harvey for the COVID-19 Pandemic Response.," *Journal of Information Technology* (35:3), pp. 195–213. (https://doi.org/10.1177/0268396220929258).

Mirbabaie, M., and Fromm, J. 2019. "Reducing the Cognitive Load of Decision-Makers in Emergency Management through Augmented Reality," in *Proceedings of the 27th European Conference on Information Systems*, pp. 1–11.

Mirbabaie, M., Stieglitz, S., and Brünker, F. 2021. "Dynamics of Convergence Behaviour in Social Media Crisis Communication – a Complexity Perspective," *Information Technology and People* (ahead-of-print), pp. 1–27. (https://doi.org/10.1108/ITP-10-2019-0537).

NATO. 2017. "Digital Hydra: Security Implications of False Information Online," Riga.

Peffers, K., Tuunanen, T., Rothenberger, M. A., and Chatterjee, S. 2007. "A Design Science Research Methodology for Information Systems Research," *Journal of Management Information Systems* (24:3). (https://doi.org/10.2753/MIS0742-1222240302).

Rathore, A. K., Maurya, D., and Srivastava, A. K. 2021. "Do Policymakers Use Social Media for Policy Design? A Twitter Analytics Approach," *Australasian Journal of Information Systems* (25:1), pp. 1–31. (https://doi.org/10.3127/ajis.v25i0.2965).

Reuter, C., and Kaufhold, M. A. 2018. "Fifteen Years of Social Media in Emergencies: A Retrospective Review and Future Directions for Crisis Informatics," *Journal of Contingencies and Crisis Management* (26:1), pp. 41–57. (https://doi.org/10.1111/1468-5973.12196).

Shahi, G. K., Dirkson, A., and Majchrzak, T. A. 2021. "An Exploratory Study of COVID-19 Misinformation on Twitter," *Online Social Networks and Media* (22:3), pp. 1–16.

Stieglitz, S., Dang-Xuan, L., Bruns, A., and Neuberger, C. 2014. "Social Media Analytics - An Interdisciplinary Approach and Its Implications for Information Systems," *Business and Information Systems Engineering* (6:2), pp. 89–96.

Stieglitz, S., Mirbabaie, M., Ross, B., and Neuberger, C. 2018. "Social Media Analytics – Challenges in Topic Discovery, Data Collection, and Data Preparation," *International Journal of Information Management* (39), pp. 156–168.

Wardle, C., and Derakhshan, H. 2017. "Information Disorder: Interdisciplinary Framework for Research and Policy," Strasbourg.

## Acknowledgements

This project has received funding from the European Union's Horizon 2020 research and innovation programme under the Marie Skłodowska-Curie grant agreement No 823866.